\documentclass[reprint, amsmath, amssymb, aps, groupedaddress, superscriptaddress]{revtex4-2}

\usepackage{graphicx}
\usepackage{dcolumn}
\usepackage{xcolor}
\usepackage{bm}
\usepackage{xspace}
\usepackage{esint}

\newcommand{\alphas}{\alpha_{\mathrm{s}}}

\newcommand{\eg}{\textit{e.g.}\xspace}
\newcommand{\ie}{\textit{i.e.}\xspace}

\newcommand{\ghost}{shadow\xspace}
\newcommand{\Cqcoll}{C^q_{\textrm{coll}}}

\begin{document}

\preprint{APS/123-QED}

\title{The deconvolution problem of deeply virtual Compton scattering}

\date{\today}

\author{V.~Bertone}
\email{valerio.bertone@cea.fr}
\affiliation{IRFU, CEA, Université Paris-Saclay, F-91191 Gif-sur-Yvette, France}
\author{H.~Dutrieux}
\email{herve.dutrieux@cea.fr}
\affiliation{IRFU, CEA, Université Paris-Saclay, F-91191 Gif-sur-Yvette, France}
\author{C.~Mezrag}
\email{cedric.mezrag@cea.fr}
\affiliation{IRFU, CEA, Université Paris-Saclay, F-91191 Gif-sur-Yvette, France}
\author{H.~Moutarde}
\email{herve.moutarde@cea.fr}
\affiliation{IRFU, CEA, Université Paris-Saclay, F-91191 Gif-sur-Yvette, France}
\author{P.~Sznajder}
\email{pawel.sznajder@ncbj.gov.pl}
\affiliation{National Centre for Nuclear Research (NCBJ), Pasteura 7, 02-093 Warsaw, Poland}

\date{\today}

\begin{abstract}
Generalised parton distributions are instrumental to study both the three-dimensional structure and the energy-momentum tensor of the nucleon, and motivate numerous experimental programmes involving hard exclusive measurements. Based on a next-to-leading order analysis and a careful study of evolution effects, we exhibit non-trivial generalised parton distributions with arbitrarily small imprints on deeply virtual Compton scattering observables. This means that in practice the reconstruction of generalised parton distributions from measurements, known as the deconvolution problem, does not possess a unique solution for this channel. In this Letter we discuss the consequences on the extraction of generalised parton distributions from data and advocate for a multi-channel analysis. 
\end{abstract}


\maketitle


Generalised parton distributions (GPDs) offer a multidimensional information on the structure of hadrons encompassing the usual parton distribution functions (PDFs) and hadron elastic form factors (EFFs). They were introduced in Refs. \cite{Mueller:1998fv, Ji:1996ek, Ji:1996nm, Radyushkin:1996ru, Radyushkin:1997ki} and have been recognised as a key element of current and future experiments, in particular at the US electron-ion collider (EIC) \cite{Accardi:2012qut, AbdulKhalek:2021gbh}, Chinese electron-ion collider (EIcC) \cite{Anderle:2021wcy}, and large hadron-electron collider (LHeC) \cite{AbelleiraFernandez:2012cc}. GPDs are used to describe a wide class of hard exclusive reactions. Among them  deeply virtual Compton scattering (DVCS) is frequently considered as the golden channel to extract GPDs thanks to its clean theoretical interpretation and the wealth of data already available for fits \cite{Moutarde:2018kwr, Moutarde:2019tqa, Cuic:2020iwt}. However, the DVCS cross-section only indirectly depends on GPDs: it is instead parameterised in terms of Compton form factors (CFFs). Factorisation theorems \cite{Radyushkin:1997ki, Collins:1998be, Ji:1998xh} allow to express CFFs as convolutions of GPDs with coefficient functions calculable at any order of perturbative QCD (pQCD). Retrieving GPDs from CFFs is therefore a major question of the field, known as the \emph{deconvolution problem}. 

Although this problem had been discussed in the early days of GPD physics, it mostly disappeared from the literature after 2003 while remaining unsolved. With the exception of the detailed status of Ref.~\cite{Diehl:2003ny}, the reviews of the field either briefly mentioned the absence of a generic deconvolution procedure \cite{Belitsky:2005qn, Boffi:2007yc, Kumericki:2016ehc} or directly asserted that a parametric form had to be used in a fit to experimental data \cite{Ji:1998pc, Goeke:2001tz, Guidal:2013rya, Mueller:2014hsa}. Elaborating on the discussion of Ref.~\cite{Freund:1999xf}, it was commonly believed that evolution would play a crucial role in a potential deconvolution procedure. However, no full-fledged theoretical argument or phenomenological feasibility proof has been put forward so far.

Even if lattice QCD offers the long-term promise of a first-principle evaluation of the functional shape of GPDs (see \eg \cite{Constantinou:2020pek}), experimental data are, and will remain for years, the main source of quantitative knowledge about GPDs. This Letter deals with the question of extracting leading twist GPDs from DVCS experimental data beyond leading order (LO). Our argument involves Lorentz covariance,  QCD evolution equations and the implementation of additional knowledge coming from PDFs and EFFs. 
We provide the first proof that the reconstruction of GPDs from CFFs using next-to-leading order (NLO) coefficient functions is ambiguous, even if the scale dependence is taken into account.

In the first section we remind the GPD description of the DVCS channel. Then, we recover the known solution of the deconvolution problem at LO before addressing its NLO variant and the role of QCD evolution. Finally, we outline the consequences for GPD phenomenology.


\section{Generalised parton distributions and DVCS amplitudes}
\label{sec:problem-statement}

GPDs are real functions of the average longitudinal light-front momentum fraction of the active parton $x$, the longitudinal momentum transfer (also called skewness) $\xi$, and four-momentum transfer to the hadron target $t$. The latter acts as a mere parameter in our study and thus will be dropped.
GPDs additionally depend on a factorisation scale $\mu$. In the following, we use the labels $q$ for quark flavours and $g$ for gluons. In view of the description of DVCS, we focus on the leading twist chiral-even singlet quark GPD $H^{q(+)}(x, \xi, \mu^2) = H^{q}(x, \xi, \mu^2) - H^{q}(-x, \xi, \mu^2)$. We specifically study the GPD $H$ for the sake of simplicity, but the discussion can be readily extended to the GPDs $E$, $\widetilde{H}$ and $\widetilde{E}$.

In the forward limit $H^{q(+)}(x,\xi,\mu^2)$ reduces to the usual PDF $f^{q(+)}(x, \mu^2)$. EFFs appear through the $0$-th order Mellin moments of GPDs in the variable $x$, hence bringing no constraint on $H^{q(+)}$ contrary to odd-order Mellin moments. Owing to Lorentz covariance, the \emph{polynomiality} property \cite{Ji:1998pc,Radyushkin:1998bz} states that the Mellin moment of order $n$ in $x$ of a quark GPD is a polynomial of order $n+1$ in $\xi$
\begin{equation}
    \int_{-1}^{1} \mathrm{d}x\,x^n H^{q(+)}(x,\xi,\mu^2) = \sum_{\substack{k = 0 \\ \mathrm{even}}}^{n+1} H^{q,n}_k(\mu^2)\, \xi^k \,.
    \label{eq:pol}
\end{equation}
It is equivalent \cite{Chouika:2017dhe, Chouika:2017rzs} for $H^{q(+)}$ to obey Eq.~\eqref{eq:pol} and to be written as the Radon transform of the sum of a double distribution (DD) $F^{q(+)}$ and a function $D^q$ called the D-term
\begin{align}
    H^{q(+)}(x,\xi,\mu^2) = \int \mathrm{d}\Omega [F^{q(+)}(\beta, \alpha, \mu^2) + \xi \delta(\beta)\,D^q(\alpha,\mu^2)]\,,
    \label{eq:radon}
\end{align}
with $\mathrm{d}\Omega =\mathrm{d}\beta\,\mathrm{d}\alpha\,\delta(x-\beta-\alpha\xi)$, where $|\alpha|+|\beta|\leq 1$.

The quantitative description of DVCS involves the CFF $\mathcal{H}$ which we assume to be known since several global determinations of CFFs from DVCS experimental data are now publicly available \cite{Moutarde:2018kwr, Moutarde:2019tqa}. $\mathcal{H}$ is the sum of scheme-dependent quark $\mathcal{H}^q$ and gluon $\mathcal{H}^g$ contributions. For the sake of simplicity we further assume that $\mathcal{H}^q$ is known. It results from the convolution of the GPD $H^{q(+)}$ and a coefficient function $T^q$
\begin{equation}
    \mathcal{H}^q(\xi, Q^2) = \int_{-1}^{1} \frac{\mathrm{d}x}{2\xi} T^q\left(\frac{x}{\xi}, \frac{Q^2}{\mu^2}, \alphas(\mu^2)\right)H^{q(+)}(x,\xi,\mu^2) \,,
    \label{eq:convol}
\end{equation}
where $Q^2$ is the virtuality of the photon mediating the interaction between the lepton beam and the proton target, and $\alphas(\mu^2)$ is the strong running coupling at scale $\mu^2\simeq Q^2$. $T^q$ can be computed in pQCD and is known for vector operators up to next-to-next-to-leading order (NNLO) \cite{Braun:2020yib}. Formulae for $T^q$ at NLO, which are at the core of the present study, can be found \eg in Refs.~\cite{Belitsky:1999sg, Moutarde:2013qs}.

Extracting GPDs from CFFs requires to \emph{deconvolute} Eq. \eqref{eq:convol}. A superficial counting of the degrees of freedom would suggest that no unique answer is to be expected since one variable is lost in the computation of the CFF: the dependence on $x$ and $\xi$ of the GPD is traded for the sole dependence on $\xi$ of the CFF. However, the problem is subtler because the coefficient function $T^q$ depends on the ratio $x/\xi$ and not on $x$ and $\xi$ separately on the one hand, and the dependencies on $x$ and $\xi$ of the GPD are tied by polynomiality on the other hand. 

Furthermore, there have been hopes that GPD evolution equations would offer a workaround to superficial counting \cite{Freund:1999xf}. These equations can be written in a generic form with a kernel $K$ computed in pQCD,
\begin{equation}
    \frac{\partial H(x,\xi,\mu^2)}{\partial\log \mu^2} = \int_{-1}^1 \frac{\mathrm{d}y}{\xi} K\left(\frac{x}{\xi},\frac{y}{\xi}, \alphas(\mu^2)\right) H(y,\xi,\mu^2) \;,
    \label{eq:evol}
\end{equation}
and entangle $x$, $\xi$ and $\mu^2$. The idea of evolving GPDs to solve the deconvolution problem originated from decades of experience in successful PDF phenomenology. However, the problem is far more complex for GPDs because of their multidimensionality.

Hereafter we use the compact notation $\otimes$ for an integral over $x \in [-1, +1]$ as \eg $\mathcal{H}^q(\xi, Q^2) = T^q \otimes H^{q(+)}$ in Eq.~\eqref{eq:convol}. Following Ref.~\cite{Moutarde:2013qs} the NLO coefficient function writes
\begin{equation}
\label{eq:nlo-coefficient-function}
    T^q = C^q_0 + \alphas(\mu^2)\,C^q_1 + \alphas(\mu^2) \log\left(\frac{Q^2}{\mu^2}\right) \Cqcoll \;,
\end{equation}
where $C^q_0$, $C^q_1$ and $\Cqcoll$ are known functions of $x$ and $\xi$.


\section{\ghost generalised parton distributions at Born order}
\label{ghost-gpd-lo}

With $\fint$ the Cauchy principal value and $e_q$ the quark fractional electric charge, the LO CFF $\mathcal{H}^q$ reads
\begin{align}
    \frac{\mathcal{H}^q(\xi, Q^2)}{e^2_q}
    &= i\pi H^{q(+)}(\xi, \xi, \mu^2) - \fint_{-1}^1 \mathrm{d}x \frac{H^{q(+)}(x, \xi, \mu^2)}{x+\xi} \;.
    \label{eq:expllo}
\end{align}
The once-subtracted LO dispersion relation,
\begin{align}
    \mathrm{Re}\,\mathcal{H}^q(\xi, Q^2) = - \mathcal{C}^q_H(Q^2)  \nonumber \\
    + \frac{1}{\pi} \fint_0^1 \mathrm{d}\xi'\,\mathrm{Im}\,\mathcal{H}^q(\xi', Q^2)
    \left(\frac{1}{\xi'-\xi}+\frac{1}{\xi'+\xi}\right)\,,
    \label{eq:subtraction}
\end{align}
connects its real and imaginary parts and involves the subtraction constant $\mathcal{C}^q_H(Q^2)$. Polynomiality ensures that the latter is an integral of the D-term. Therefore, if the D-term is omitted, the LO CFF is null if and only if its imaginary part is null, \ie if the GPD possesses a null diagonal $x=\xi$. A LO analysis of DVCS is therefore impervious to such distributions. To make them fully invisible to experimental DVCS and DIS data, we still need to cancel their forward limit. We call \emph{\ghost GPD} a GPD with a null CFF and a null forward limit at a given scale $\mu^2$.
The generic existence of such \ghost GPDs at finite order of pQCD is the main finding of this Letter. By linearity of Eq.~\eqref{eq:convol} \ghost GPDs at a specific scale populate a vector space, meaning that any multiple of such GPDs can be introduced in an analysis of DVCS data with no effect on theoretical predictions at that scale.

Since polynomiality plays a key role in this analysis, and is best expressed in terms of DDs, we similarly pair \emph{\ghost DDs} with \ghost GPDs. Since any continuous DD can be uniformly approximated by polynomials with an arbitrary precision, we restrict our investigations to polynomial of total degree $N$ in the variables $\alpha$ and $\beta$ 
\begin{equation}
\label{eq:def-polynomial-dd}
    F^{q(+)}(\beta, \alpha) = \sum_{m\tiny{\textrm{\ even}}, n\tiny{\textrm{\ odd}}}^{m+n \leq N} c_{mn} \, \alpha^m\beta^n\,.
\end{equation}
Its definition requires $(N+1)(N+3)/8$ coefficients $c_{mn}$. It parameterises a GPD that writes for $x > |\xi|$ as 
\begin{equation}
\label{eq:gpd-from-polynomial-dd}
    H^{q(+)}(x, \xi) = \sum_{u=1, v=0}^{N+1} \left[\frac{1}{(1+\xi)^u} + \frac{1}{(1-\xi)^u}\right] q_{uv} \, x^v\,,
\end{equation}
with a \emph{linear relation} between coefficients $q_{uv}$ and $c_{mn}$
\begin{equation}
\label{eq:radon-coefficients}
    q_{uv} = \sum_{m, n} R_{uv}^{mn}\,c_{mn}\,,
\end{equation}
inherited from the Radon transform
\begin{align}
    R^{mn}_{uv} = \sum_{j=0}^n \frac{(-1)^{u+v+j}}{m+j+1} \begin{pmatrix}n \\ j\end{pmatrix}\begin{pmatrix}j  \\ m-u+j+1\end{pmatrix} \nonumber \\\times \begin{pmatrix}m+j+1 \\ v-n+j\end{pmatrix}\,.
\end{align}
The concatenation of the coefficients $c_{mn}$ and $q_{uv}$ in two vectors recasts Eq.~\eqref{eq:radon-coefficients} as the action of a matrix denoted by $R$. 

To build a \ghost GPD with the DD \eqref{eq:def-polynomial-dd} we have to enforce both $H^{q(+)}(\xi, \xi) = 0$ and $H^{q(+)}(x, 0) =0$. We notice that for $\xi \geq 0$
\begin{eqnarray}
    H^{q(+)}(\xi, \xi) & = & \sum_{w = 1}^{N+1} \frac{1}{(1+\xi)^w} \sum_{u,v} C^{uv}_w \, q_{uv} \,, \label{eq:diag} \\
    H^{q(+)}(x, 0) & = & \sum_{w=0}^{N+1} x^w \sum_{u,v} Q^{uv}_w \, q_{uv} \;,
\end{eqnarray}
where the matrices $C$ and $Q$ have elements
\begin{eqnarray}
    C^{uv}_w = (-1)^{u+v+w}\begin{pmatrix}v \\ u-w\end{pmatrix},\quad \quad 
    Q^{uv}_w = 2 \delta^{v}_w \,,
\end{eqnarray}
where $\delta^v_w$ is the Kronecker symbol. Finding a LO \ghost DD amounts to solving the systems $C\cdot R = 0$ and $Q \cdot R = 0$. They respectively contain $N+1$ and $N+2$ equations for $(N+1)(N+3)/8$ unknown coefficients $c_{mn}$. They result in an \emph{underconstrained} problem which admits a vector space of solutions of dimension growing quadratically with $N$. 
An explicit example of an infinite family of LO \ghost DDs is provided in App.~\ref{appendix:infinite-family-lo-shadow-dd}.


\section{\ghost generalised parton distributions at one loop}
\label{ghost-gpd-nlo}

The existence of infinitely many GPDs with a null LO CFF at a given scale was already known. However, to the best of our knowledge the vanishing of the LO CFF has never been studied before in conjunction with the vanishing of the forward limit of the GPD. Moreover, the advantage of the analysis of the previous section is that it can be readily extended to higher orders in pQCD. The NLO CFF can be naturally split into three parts corresponding to the structure of the NLO coefficient function \eqref{eq:nlo-coefficient-function}. As seen above for a polynomial DD of degree $N$, the number of free parameters scales as $N^2$, while the number of constraints brought by the cancellation of the LO CFF or the forward limit scales as $N$. For $N$ large enough we expect to find infinitely many solutions. Our strategy at NLO consists in cancelling in a similar way the two additional terms $C^q_1 \otimes H^{q}$ and $\Cqcoll \otimes H^{q}$. Thanks to dispersion relations \cite{Diehl:2007jb} and the omission of the D-term, it is enough to compute the imaginary parts of these two terms. 

The contribution to $\mathrm{Im}\, \mathcal{H}^q$ originating from $\Cqcoll$ yields

\begin{align}
\label{eq:integral-in-cff-collinear-part}
    \mathrm{Im}\,\Cqcoll \otimes H^q = \frac{e^2_q C_F}{2} \bigg(H^{q(+)}(\xi, \xi) \left[\frac{3}{2} + \log \left(\frac{1-\xi}{2\xi}\right)\right] \nonumber \\
    + \int_\xi^1 \mathrm{d}x\,\frac{H^{q(+)}(x,\xi)-H^{q(+)}(\xi,\xi)}{x-\xi}\bigg)\,,
\end{align}
where $C_F = 4/3$. For the particular case of a polynomial DD, $H^{q(+)}(x,\xi)$ is a polynomial in $x$ (see Eq.~\eqref{eq:gpd-from-polynomial-dd}) and the integral is well-defined. It can be explicitly evaluated

\begin{align}
\label{eq:integral-in-cff-collinear-part-2}
\int_\xi^1 \mathrm{d}x\,\frac{H^{q(+)}(x,\xi)-H^{q(+)}(\xi,\xi)}{x-\xi} = \sum_{w=1}^{N+1} \frac{\sum_{u,v} D^{uv}_w\, q_{uv} }{(1+\xi)^w} \;,
\end{align}

where
\begin{align}
    D^{uv}_w = (-1)^{u+v+w}\sum_{k=1}^{v} \frac{(-1)^k}{k} \begin{pmatrix}v-k \\ u-w\end{pmatrix}-\frac{1}{k}\begin{pmatrix}v \\ u-w\end{pmatrix} \,.
    \label{eq:systemD}
\end{align}
The term coming with a logarithmic singularity in Eq.~\eqref{eq:integral-in-cff-collinear-part} disappears by requiring $\mathrm{Im}\,C^q_0 \otimes H^{q} = 0$. 
Enforcing $\mathrm{Im}\,\Cqcoll \otimes H^q = 0$ thus brings at most $N+1$ extra linear conditions. 
We similarly cast these constraints in a matrix form $D \cdot R = 0$.

We proceed along the same lines with $\mathrm{Im}\,C^q_1 \otimes H^q$ and remarkably find that if $\mathrm{Im}\,C^q_0 \otimes H^{q} = 0$,
\begin{align}
    \mathrm{Im}\,C^q_1 \otimes H^q = \log\left(\frac{1-\xi}{2\xi}\right)\mathrm{Im}\,C^q_{coll} \otimes H^q \nonumber \\ +  \frac{e^2_q C_F}{4}\sum_{w=1}^{N-1}\frac{1}{(1+\xi)^w} \sum_{u,v} E^{uv}_w \, q_{uv} \;,
\end{align}
where
\begin{align}
    E^{uv}_w = (-1)^{u+w}  \sum_{k=1}^{v} \frac{3}{k} \begin{pmatrix}v-k \\ u-w\end{pmatrix}-\frac{3(-1)^k}{k}\begin{pmatrix}v \\ u-w\end{pmatrix}\nonumber\\ -  \frac{2(-1)^v}{k}\sum_{j=1}^k \frac{(-1)^j}{j}\begin{pmatrix}v-j \\ u-w\end{pmatrix} - \frac{1}{j}\begin{pmatrix}v \\ u-w\end{pmatrix}\,.
    \label{eq:systemE}
\end{align}
Here, again the extra cancellation of $\mathrm{Im}\,C^q_1 \otimes H^q$ invokes at most $N-1$ new equations displayed in matrix form $E \cdot R = 0$.

We also note that a GPD resulting from a DD $F^{q(+)}$ may exhibit a discontinuity at $(x, \xi) = (1, 1)$. For instance for any $\lambda \geq 1$,
\begin{equation}
    \lim_{\varepsilon\rightarrow 0} H^{q(+)}\left(1-\frac{\varepsilon}{\lambda}, 1-\varepsilon\right) = \int_{0}^{1/\lambda} \mathrm{d}\alpha\,F^{q(+)}\left(1-\alpha, \alpha\right) \,,
\end{equation}
and the limit at $(x, \xi) = (1, 1)$ may depend on the actual path to $(1, 1)$ in the $(x, \xi)$-plane, unless $F^{q(+)}(1-\alpha, \alpha) = 0$. This adds another set of equations $B \cdot R = 0$ with
\begin{equation}
    F^{q(+)}(1-\alpha, \alpha) = \sum_{w=0}^N \alpha^w \, \sum_{m, n} B^{mn}_w \, c_{mn} \;,
\end{equation}
where
\begin{equation}
    B^{mn}_w = (-1)^w \begin{pmatrix} n \\ w-m \end{pmatrix}\,.
\end{equation}

\begin{figure}
\includegraphics[]{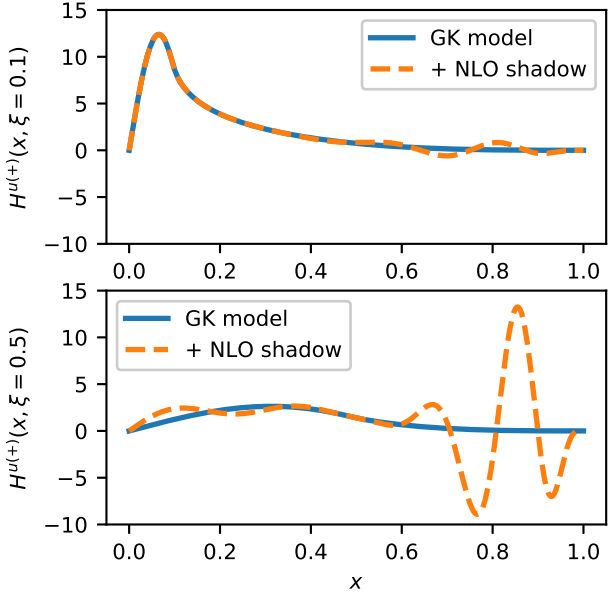}
\caption{\label{fig:gpdnonlo} $H^{u(+)}$ as a function of $x$ for $\xi = 0.1$ and 0.5, $t = -0.1$ GeV$^2$ and $\mu_0^2 = 1$ GeV$^2$. Solid blue: Goloskokov-Kroll (GK) model \cite{Goloskokov:2005sd, Goloskokov:2007nt, Goloskokov:2009ia}. Dashed orange: GK model with the addition of an NLO \ghost GPD. In both cases one obtains exactly the same NLO CFF and forward limit at scale $\mu_0^2$.}
\end{figure}

An example of such NLO \ghost quark GPD satisfying simultaneously $Q \cdot R = 0$ (forward limit), $C \cdot R = 0$ ($C_0$), $D \cdot R = 0$ ($\Cqcoll$), $E \cdot R = 0$ ($C_1$) and $B \cdot R = 0$ (continuity) is given in Fig.~\ref{fig:gpdnonlo}, where it is added to a popular phenomenological GPD model. Although $F^{q(+)}$ is a polynomial of degree 25, the \ghost GPD does not oscillate in a way \textit{a priori} excluded on physical grounds. We foresee that the argument of the relative increase of the number of constraints and of free parameters can be extended \textit{mutatis mutandis} to guarantee the existence of \ghost gluon GPDs and more generally of \ghost GPDs at any finite order in pQCD. Similarly the knowledge of the first few Mellin moments \eqref{eq:pol} would only bring a finite number of new constraints within the scope of this analysis and would not alter our conclusions. 

We have so far only worked at a given scale without considering the effect of evolution. However, in the phenomenology of parton distributions a model is defined at an arbitrary scale $\mu_0^2$ and then evolved to a scale $\mu^2 \simeq Q^2$, where the theoretical prediction is compared to experimental data. Therefore, we now consider the case where a \ghost quark GPD $H^{q(+)}$ is added to a phenomenologically relevant model at the scale $\mu_0^2$. Thanks to the linearity of both the CFF convolution \eqref{eq:convol} and evolution equations \eqref{eq:evol}, we can compute the specific CFF contribution of this \ghost GPD while all other GPDs are taken to be 0 at $\mu_0^2$.

In terms of the kernel $K_{ab}$ for parton types $a, b \in \{q, g\}$, the GPD evolution operator $\Gamma_{ab}(\mu^2, \mu_0^2)$ between scales $\mu_0^2$ and $\mu^2$ obtained by solving the evolution equation \eqref{eq:evol} admits the expansion
\begin{align}
\label{eq:evolution-operator-alphas}
    \Gamma_{ab}(\mu^2, \mu_0^2) &= 1 + \alphas(\mu^2) K^{(0)}_{ab} \log\left(\frac{\mu^2}{\mu_0^2}\right) + \mathcal{O}(\alphas^2(\mu^2)) \;,
\end{align}
provided that $\mu^2$ is close enough to $\mu_0^2$. Being observable, a CFF cannot depend on $\mu^2$. At one loop this yields
\begin{align}
    \Cqcoll + C^q_0 \otimes K^{(0)}_{qq} = 0\,.
\end{align}
In particular, $C^q_0 \otimes K^{(0)}_{qq} \otimes H^{q(+)}(\mu_0^2) = 0$ for the NLO \ghost quark GPD.
Its contribution to the quark component of the NLO CFF becomes
\begin{align}
\label{eq:pqcd-consistency-check}
    \mathcal{H}^q(\xi, Q^2) = C^q_0 \otimes H^{q(+)}(\mu_0^2) + \alphas(\mu^2)\, C^q_1 \otimes H^{q(+)}(\mu_0^2) \nonumber \\ + \alphas(\mu^2)\,C^q_0 \otimes K^{(0)}_{qq} \otimes H^{q(+)}(\mu_0^2) \log \left(\frac{\mu^2}{\mu_0^2}\right) \nonumber \\ + \alphas(\mu^2)\, \Cqcoll \otimes H^{q(+)}(\mu_0^2) \log\left(\frac{\mu^2}{Q^2} \right)  + \mathcal{O}(\alphas^2(\mu^2)) \,.
\end{align}
By definition of NLO \ghost quark GPDs, the term $\propto\,\alphas^0(\mu^2)$ and all three terms $\propto\,\alphas^1(\mu^2)$ cancel, resulting in  $\mathcal{H}^q(\xi, Q^2) = \mathcal{O}(\alphas^2(\mu^2))$. Since gluons enter DVCS at $\mathcal{O}(\alphas^1(\mu^2))$ and are radiatively generated as $\mathcal{O}(\alphas^1(\mu^2))$, the conclusion is not limited to the quark sector.

\begin{figure}
\includegraphics[]{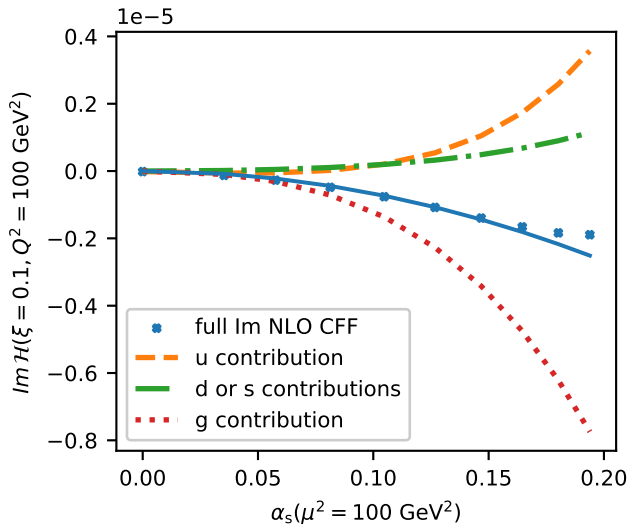}
\caption{\label{fig:cffnonlo} Imaginary part of the NLO CFF $\mathcal{H}(\xi = 0.1, Q^2 = 100~\mathrm{GeV}^2)$ evaluated with the NLO \ghost GPD shown in Fig.~\ref{fig:gpdnonlo} and $H^{d(+)} = H^{s(+)} = H^{g(+)} = 0$ at $\mu_0^2 = 1$ GeV$^2$. The blue dots correspond to computations made with different values of $\alphas(\mu^2 = 100~\mathrm{GeV}^2)$ and the solid blue line results from a quadratic fit to the first seven points. The dashed orange, dash-dotted green and dotted brown lines indicate $u$, $d$ or $s$, and $g$ contributions to the CFF, respectively.}
\end{figure}

This pQCD prediction can be probed in a realistic setting $\xi = 0.1$ and $Q^2 = 100~\mathrm{GeV}^2$ relevant for future colliders. We consider the NLO \ghost $H^{u(+)}$ of Fig.~\ref{fig:gpdnonlo} at $\mu_0^2 = 1~\mathrm{GeV}^2$ and evolve it to $\mu^2 = 100~\mathrm{GeV}^2$, while dynamically generating $d$, $s$ and $g$ GPDs by evolution. Using APFEL++ \cite{Bertone:2013vaa,Bertone:2017gds,Bertone:2021} and PARTONS \cite{Berthou:2015oaw} we compute the corresponding NLO CFF and continuously change the value of $\alphas(\mu^2)$. Fig.~\ref{fig:cffnonlo} shows that as expected (i) $\mathrm{Im}\, \mathcal{H} = \mathcal{O}(\alphas^2(\mu^2))$ and (ii) $\mathrm{Im}\, \mathcal{H} \lesssim 10^{-5}$ for $\xi = 0.1$ and $Q^2 = 100~\mathrm{GeV}^2$. It means that the CFF contribution is smaller than the \ghost GPD at $\xi = 0.1$ by about 5 orders of magnitude. In particular such a \ghost GPD will be hidden in the typical statistical and systematic uncertainties of DVCS measurements. Contradicting many claims as old as Ref.~\cite{Freund:1999xf}, we have proved here by a theoretical argument and an explicit quantitative example that evolution alone will not solve the deconvolution problem.

\section{Discussion}
\label{sec:discussion}

We have shown the existence of \ghost GPDs \ie nontrivial distributions with a null forward limit and negligible contributions to CFFs over kinematic domains relevant for the experiments at Jefferson Lab, CERN, EIC or EIcC. 
This result holds in a full next-to-leading order analysis of DVCS and crucially relies on properties grounded in first principles as analyticity, polynomiality and evolution. 
We expect it to remain true at any finite perturbative order. 
Our result is the first quantitative answer to the deconvolution problem beyond leading order and it points to a non-uniqueness of the solution.

Since experimental data are acquired at non-zero skewness $\xi$, the extrapolation towards zero skewness required for proton tomography may suffer from the residual presence of \ghost distributions. 
The experimental tomography of the singlet sector or the determination of the parton orbital angular momentum will require extra care, especially since the  forward limit of the GPD $E$ is only weakly constrained, contrary to the GPD $H$. 
Had we not omitted the D-term in this discussion, a \ghost GPD would originate from the sum of a \ghost double distribution and a D-term.
A detailed investigation of this extension of the concept of a \ghost GPD would be interesting.
In particular the first Mellin moment of the D-term is instrumental to the extraction of proton mechanical properties, which is highly sensitive to the uncertainties associated to existing data \cite{Dutrieux:2021nlz}.

Measurements of Compton scattering (both timelike and spacelike), meson production (at least at leading order) or the lattice QCD evaluation of the first few Mellin moments are not expected to qualitatively modify this answer to the deconvolution problem. In terms of fits to measurements, only multi-channel analysis of experimental data beyond leading order, over wide kinematic domains accessible in collider experiments, and within a complete framework such as PARTONS may provide the needed leverage to quantitatively constrain GPDs. Double DVCS offers a direct access to GPDs at $x\neq \xi$ and seems a natural candidate to make the deconvolution well-defined. It may also be desirable to constrain GPDs extracted from experimental data with lattice QCD computations in $x$-space or with additional physical principles. In particular, the positivity property will bound the range of variation of \ghost GPDs, although this bound will be weak when $x \simeq \xi$ because of the divergence of phenomenological proton PDFs at small $x$. The generic existence of \ghost distributions therefore calls for many subsequent studies of the phenomenology of generalised parton distributions.

\begin{acknowledgments}
The authors thank B.~Pire and J.~Wagner for fruitful discussions. This project was supported by the European Union's Horizon 2020 research and innovation programme under grant agreement No 824093. This work was supported by the Grant No. 2019/35/D/ST2/00272 of the National Science Center, Poland. The project is co-financed by the Polish National Agency for Academic Exchange and by the COPIN-IN2P3 Agreement. 
\end{acknowledgments}

\appendix
\section{An example of LO \ghost double distributions}
\label{appendix:infinite-family-lo-shadow-dd}

An infinite family of LO \ghost DDs is made of the following polynomials of odd order $N \geq 9$
\begin{widetext}
\begin{align}
F^{q(+)}_{N}(\beta, \alpha) = \beta^{N-8}\bigg[\alpha^8-\frac{28}{9}\alpha^6\bigg(\frac{N^2-3N+20}{(N+1)N}+\beta^2\bigg)+\frac{10}{3}\alpha^4\bigg(\frac{N^2-7N+40}{(N+1)N}+\frac{2(N^2-3N+44)}{3(N+1)N}\beta^2+\beta^4\bigg)\nonumber\\
-\frac{4}{3}\alpha^2\bigg(\frac{N^2-11N+60}{(N+1)N}-\frac{N-8}{N}\beta^2-\frac{N^2-3N-28}{(N+1)N}\beta^4+\beta^6\bigg)+\frac{1}{9}(1-\beta^2)^2\bigg(\frac{N^2-15N+80}{(N+1)N}-\frac{2(N-8)}{N}\beta^2+\beta^4\bigg)\bigg]\,.
\end{align}
\end{widetext}

\section{Open source code}
\label{appendix:opensource_code}

The analytic form of the \ghost GPD displayed in Fig.~\ref{fig:gpdnonlo} is available in the PARTONS framework \cite{Berthou:2015oaw} as the module \texttt{GPDBDMMS21}. The code of this framework is open source and can be found online at \url{https://drf-gitlab.cea.fr/partons/core/partons} on version 3 of the GPL (GPLv3).

\bibliography{deconvolution}

\end{document}